# MIMO Precoding Using Rotating Codebooks

*C Jiang, M Wang, C Yang*

*Abstract*—Next generation wireless communications rely on multiple input multiple output (MIMO) techniques to achieve high data rates. Feedback of channel information can be used in MIMO precoding to fully activate the strongest channel modes and improve MIMO performance. Unfortunately, the bandwidth of the control channel via which the feedback is conveyed is severely limited. An important issue is how to improve the MIMO precoding performance with minimal feedback. In this letter, we present a method that uses a rotating codebook technique to effectively improve the precoding performance without the need of increasing feedback overhead. The basic idea of the rotating codebook precoding is to expend the effective precoding codebook size via rotating multiple codebooks so that the number of feedback bits remains unchanged. Simulation results are presented to show the performance gain of the proposed rotating codebook precoding over the conventional precoding.

## I. INTRODUCTION

MULTIPLE input multiple output (MIMO) systems have become the most promising candidates for the next generation of high data rate wireless communications [1]. MIMO technology has been shown to provide significant system performance improvement over conventional systems by providing communication links with substantial diversity and capacity. This is especially true when channel state information (CSI) is available at the transmitter [2]. In fact, the closed loop capacity of a MIMO channel can be achieved by converting the channel into a set of parallel spatial layers via precoding and water-filling power allocation at the transmitter and a linear MMSE filtering at the receiver wherein the optimal precoding and MMSE filter are determined by the singular value decomposition (SVD) of the MIMO channel matrix [3][4].

For time division duplexing (TDD) communication systems, where uplink and downlink channels share the same frequency band at different time, the down link channel information can be estimated by the transmitter using the uplink pilot (analogy feedback) due to the channel reciprocity property. For frequency division duplexing (FDD) systems, where different frequency bands are allocated for the downlink and the uplink channels, the channels are not reciprocal. Therefore, a CSI feedback channel, via the uplink control channel (digital feedback), is necessary to deliver the estimated channel knowledge at the receiver back to the transmitter. For a time varying channel, the channel knowledge at the transmitter should be updated regularly. The overhead, which increases linearly as the product of the number of antennas, the channel frequency selectivity and the feedback frequency, can be large. Hence, the MIMO technique utilizing full channel information is impractical for the system with limited feedback channel capacity.

Limited feedback MIMO signaling has been extensively researched in the past [6]-[12]. In particular, unitary precoding [8]-[10] has shown to be an effective form of limited feedback MIMO that quantizes the precoder rather than the channel information. The essential idea is that the transmit precoder is chosen from a finite set of precoding matrices, called the codebook, known *a priori* to both the transmitter and receiver (the transceivers). The receiver chooses the optimal precoder (precoding matrix) from the codebook as a function of the current channel state information available at the receiver and sends the index of this matrix to the transmitter over a limited bandwidth control channel. This closed loop precoding technique provides more flexibility than the direct channel quantization scheme in the feedback limited channel and has been quickly adopted by the next generation wireless standards to improve MIMO performance in low mobility environments [13][14][15]. Although the unitary precoding technique has made the closed loop MIMO precoding more practical, the feedback overhead can still be significant depending on the number of antennas, the number of sub-bands and the feedback frequency. Therefore, the number of feedback bits for a practical system is typically limited to 4 bits per sub-band [13][14] which is far from sufficient to fully exploit the MIMO precoding gain especially for systems with large antenna arrays (e.g., eight transmission antennas [18]) and systems that exploit multi-user MIMO gains.

In this letter, a proposed technique, referred to as the rotating codebook precoding, is aimed at further exploiting the MIMO precoding gain *without* incurring increased feedback. First off, is a brief review of the unitary precoding in Section II. Secondly, the rotating codebook precoding technique and its performance simulation results are presented in Section III, followed by a conclusion in Section IV.

## II. Unitary Precoding

First, let's consider a linear precoding MIMO system where $\mathbf{s} = [s_1, s_2, \cdots, s_M]^T$ contains a vector of $M$ modulation symbols with $E\{\mathbf{ss}^*\} = \mathbf{I}_M$. The symbol vector $\mathbf{s}$ is then multiplied by an $M_T \times M$ precoding matrix $\mathbf{F}$, which defines a mapping between the $M_T$ ($M_T \geq M$) transmit antennas and the modulation symbol vector $\mathbf{s}$, producing a vector $\mathbf{x} = \mathbf{Fs}$ of length $M_T$. Assuming a narrowband MIMO channel (e.g., a subcarrier of an OFDM system) with perfect synchronization, the baseband, discrete-time equivalent received signal can be written as

$$\mathbf{y} = \mathbf{Hx} + \mathbf{w} \qquad (1)$$

where $\mathbf{H}$ is an $M_R \times M_T$ channel matrix with independent entries identically distributed according to $\mathcal{CN}(0,1)$, $M_R \geq M$ is the number of receive antennas and $\mathbf{w}$ is a length $M_R$ noise vector. Again, we assume that the entries of $\mathbf{w}$ are independent and distributed according to $\mathcal{CN}(0, N_0)$.

In closed-loop precoding, the receiver chooses an $M_T \times M$ precoding matrix $\mathbf{F}$. The transmitter precodes the symbol vector $\mathbf{s}$ with $\mathbf{F}$, conveyed from the receiver, before transmission. The equivalent channel is $\mathbf{HF}$. Thus, the capacity given MIMO channel matrix $\mathbf{H}$ and the precoding matrix $\mathbf{F}$ is

$$C(\mathbf{F}) = \log \det \left( I + \gamma (\mathbf{HF})^* (\mathbf{HF}) \right) = \log \det \left( I + \gamma \mathbf{F}^* \mathbf{H}^* \mathbf{HF} \right) \quad (2)$$

where $\gamma$ is the ratio of the transmit signal power to the receiver noise power [16]. The best precoder that maximizes the capacity is

$$\hat{\mathbf{F}} = \arg \max_{\mathbf{F}} C(\mathbf{F}) \quad (3)$$

subject to the unitary constraint, $\mathbf{F}^* \mathbf{F} = \mathbf{I}_M$. Based on the fact that function extrema are obtained when the matrix variables have the same eigenvectors [17], the optimal beam directions are given by the eigenvectors of

$$\mathbf{H}^* \mathbf{H} = \left( \mathbf{U} \mathbf{\Lambda} \mathbf{V}^* \right)^* \mathbf{U} \mathbf{\Lambda} \mathbf{V}^* = \mathbf{V} \mathbf{\Lambda}^2 \mathbf{V}^* \quad (4)$$

where $\mathbf{H}$ is decomposed via singular value decomposition $\mathbf{H} = \mathbf{U} \mathbf{\Lambda} \mathbf{V}^*$. $\mathbf{U}$ and $\mathbf{V}$ are unitary matrices and $\mathbf{\Lambda}$ is real diagonal (singular values). The unitary precoding matrix that maximizes C is given by

$$\hat{\mathbf{F}} = \mathbf{V}. \quad (5)$$

That is, the optimal unitary precoding matrix is the channel's eigen matrix. This structure is essentially a beamformer with multiple beams (equal powers) that maps or beamforms data symbols across the spatial channels in order to maximize capacity or minimize error rate.

Feeding back the actual $\hat{\mathbf{F}}$ from a receiver to the transmitter is costly, and can be practically prohibitive even if $\hat{\mathbf{F}}$ is limited to a unitary matrix. In the limited feedback unitary precoding system, as illustrated in Fig. 1, $\hat{\mathbf{F}}$ is further confined to a limited number of choices, i.e., $\hat{\mathbf{F}} \in \mathscr{F}^L$, where $\mathscr{F}^L$ is a finite set of unitary matrices of size $|\mathscr{F}^L| = 2^L$,

$$\mathscr{F}^L \triangleq \left\{ \mathbf{F}_0, \mathbf{F}_1, \cdots, \mathbf{F}_i, \cdots \mathbf{F}_{2^L - 1} \right\}, \quad (6)$$

called a codebook of precoding matrices. Obviously, the size of the codebook is limited by the number of bits per feedback, $L$. Each precoding matrix is an $M_T \times M$ unitary matrix, i.e., $\mathbf{F}_i^* \mathbf{F}_i = \mathbf{I}_M$, $0 \leq i \leq 2^L - 1$. The $2^L$ precoding matrices is independent of the current channel state and thereby can be predetermined and made available at both ends (i.e., the transceivers) of the communication link. Based on the estimated channel knowledge, the receiver selects a proper precoding matrix from the codebook based on a certain criterion, such as the one given by (3),

$$\mathbf{F}_{\hat{i}} = \arg\max_{\mathbf{F}_i \in \mathcal{F}^L} C(\mathbf{F}_i), \tag{7}$$

and then conveys the corresponding code index, i.e., the *L*-bit preferred precoding matrix index $\hat{i}$, to the transmitter via a control channel for symbol mapping characterized by the preferred precoding matrix $\mathcal{F}^L(\hat{i}) = \mathbf{F}_{\hat{i}}$.

Fig. 2 shows the closed-loop unitary precoding performance with various codebook sizes where *L*=0 corresponds to the open loop scenario where there is no feedback and precoding matrices are randomly chosen from a codebook at each transmission at the transmitter. $L = \infty$ denotes that the actual (not quantized) $\mathbf{F}$ from (5) is fed back to the transmitter. The creation of variable size unitary precoding codebooks can be found in [8]-[10]. The delay from the measurement of the channel to the application of the precoding matrix is ~3 ms. Like any other feedback scheme, the closed-loop precoding gain diminishes quickly as the channel mobility increases due to the feedback delay. It is evident that the closed-loop precoding gain over open loop is dependent on the number of feedback bits or the size of the codebook, especially in a low mobility environment. Typically, *L* is limited to 4 bits per subband per feedback to avoid incurring excessive control channel overhead [13]. Clearly, there is a significant gap between $L = 4$ and $L = \infty$ for improvement. The gap increases as the number of antennas increases. The goal of the proposed rotating codebook scheme is to minimize the gap without invoking additional feedback bits.

### III. ROTATING CODEBOOKS FOR LIMITED FEEDBACK PRECODING

It is known that the performance of the limited feedback precoding system in a low mobility environment depends on the size of the codebook, determined by the number of feedback bits. Increasing the number of feedback bits no doubt enhances the performance. However, the increase in the number of feedback bits in turn requires a significant increase in the control channel bandwidth.

The goal of our rotating codebook scheme is to improve the closed-loop precoding gain without increasing the number of feedback bits. First, *K* different but equivalent codebooks of size $2^L$ are created:

$$\left\{\mathcal{F}_k^L, 0 \leq k \leq K-1\right\}. \tag{8}$$

Each codebook in this set of codebooks is alternatively used by the transceivers in a predetermined fashion. In general, the codebook used at time *t* is determined by a predetermined sequence $k = \rho(t)$, $0 \leq \rho(t) \leq K-1$, known to both the transmitter and the receiver. The codebook used at feedback time *t* is hence $\mathcal{F}_{\rho(t)}^L$. In this paper, the codebooks simply rotate in a sequential order, $\rho(t) = n \bmod K$, at each feedback time, $t = nT$, where *T* is the feedback interval, i.e.,

$$\cdots, \mathcal{F}_0^L, \mathcal{F}_1^L, \mathcal{F}_2^L, \cdots, \mathcal{F}_{K-1}^L, \mathcal{F}_0^L, \mathcal{F}_1^L, \mathcal{F}_2^L, \cdots. \tag{9}$$

This rotating codebook scheme is illustrated in Fig. 3 where the codebook used at feedback time $t$ is $\mathcal{F}_{\rho(t)}^L$, in contrast to the conventional closed-loop precoding (c.f., Fig. 1) where a same codebook $\mathcal{F}^L$ is used at all times:

$$\cdots, \mathcal{F}^L, \mathcal{F}^L, \mathcal{F}^L, \cdots, \mathcal{F}^L, \mathcal{F}^L, \mathcal{F}^L, \mathcal{F}^L, \cdots. \tag{10}$$

This rotating codebook mechanism gives the transceivers the chance to use more precoding matrices over time than a single codebook could offer without using additional feedback bits. This approach offers the transceivers a $K$ times larger virtual codebook, $\mathcal{F}^{L+\log K}$, but still using $L$ feedback bits instead of the usual $L+\log K$ bits for this size of codebook. However, since only one codebook can be used at a particular feedback time $t$, i.e., the matrix selection process is performed on one codebook, $\mathcal{F}_{\rho(t)}^L$, of size $2^L$ (other than $K2^L$),

$$\mathbf{F}_{\hat{i}} = \arg\max_{\mathbf{F}_i \in \mathcal{F}_{\rho(t)}^L} C(\mathbf{F}_i), \tag{11}$$

and since each individual codebook $\mathcal{F}_{\rho(t)}^L$, $\forall t$ is equivalent to the codebook $\mathcal{F}^L$ in (7) with the same size $2^L$, the rotating codebook scheme in this form provides no overall performance advantage over the conventional precoding that uses a single codebook with the same number of feedback bits.

In order to benefit from the rotating codebooks, the *default precoding matrix* scheme must be employed. The default precoding matrix scheme reserves an index $\tilde{i} \in \{0, \cdots, 2^L - 1\}$ for serving as an indication to the transmitter that the default precoding matrix (i.e., the precoding matrix used in the previous transmission which is a matrix optimized through the previous precoding matrix selection process) should be used for current transmission. For each codebook $\mathcal{F}_k^L$, $0 \leq k \leq K-1$, the code matrix with index $\tilde{i}$, i.e., $\mathbf{F}_{\tilde{i}} = \mathcal{F}_k^L(\tilde{i})$, is therefore punctured for use as the default matrix. That is, the original $\tilde{i}$th matrix $\mathcal{F}_k^L(\tilde{i})$ in $\mathcal{F}_k^L$ is no longer available for selection. The transceivers keep track of the precoding matrix $\tilde{\mathbf{F}}$ that is used in the previous transmission. If the receiver cannot find a precoding matrix, from the current codebook, that is better than the precoding matrix $\tilde{\mathbf{F}}$, i.e., $C(\mathbf{F}_{\hat{i}}) < C(\tilde{\mathbf{F}})$, the reserved index (i.e., the default precoding matrix indicator) $\tilde{i}$ is communicated to the transmitter. Otherwise, $\hat{i}$ is fed back to the transmitter. When the transmitter receives the reserved index $\tilde{i}$, it will continue using the previous precoding matrix, $\tilde{\mathbf{F}}$, for the current transmission. This allows the optimization results from the previous codebooks to be used in the optimization process within the current codebook. From an equivalent point of view, the current codebook $\mathcal{F}_{\rho(t)}^L$ is altered by replacing the original $\tilde{i}$th precoding matrix, $\mathcal{F}_{\rho(t)}^L(\tilde{i})$, with the previously selected precoding matrix $\tilde{\mathbf{F}}$, i.e., $\mathcal{F}_{\rho(t)}^L(\tilde{i}) = \tilde{\mathbf{F}}$. The best precoding matrix is then selected from the altered codebook, denoted as $\tilde{\mathcal{F}}_{\rho(t)}^L$, based on current channel conditions

$$\mathbf{F}_{\hat{i}} = \arg\max_{\mathbf{F}_i \in \tilde{\mathcal{F}}_{\rho(t)}^L} C(\mathbf{F}_i). \tag{12}$$

Hence, (9) becomes

$$\cdots, \tilde{\mathcal{F}}_0^L, \tilde{\mathcal{F}}_1^L, \tilde{\mathcal{F}}_2^L, \cdots, \tilde{\mathcal{F}}_{K-1}^L, \tilde{\mathcal{F}}_0^L, \tilde{\mathcal{F}}_1^L, \tilde{\mathcal{F}}_2^L, \cdots. \tag{13}$$

Since $\tilde{\mathcal{F}}_{\rho(t)}^L$ contains precoding matrices not just from the current codebook $\mathcal{F}_{\rho(t)}^L$ but also the optimal precoding matrix from the previous codebooks, (12) can be equivalently written as (ignoring the punctured matrix for now)

$$\mathbf{F}_{\hat{i}} = \arg\max_{\mathbf{F}_i \in \{\mathcal{F}_{\rho(\tau)}^L, \tau = t, t-1, \cdots\}} C(\mathbf{F}_i) \tag{14}$$

Compared to (11), it is clear that this mechanism practically enables the transceivers to choose an optimal precoding matrix that is not just optimized over the current codebook but is also optimized over previously rotated codebooks in a progressive or incremental manner. This incremental optimization process *in effect* increases the size of the codebook used at each feedback time, creating a virtual codebook that is larger than the actual one used at any feedback instance. It is thus clear that, in a static environment, the rotating codebooks $\{\mathcal{F}_k^L, 0 \leq k \leq K-1\}$ with feedback bits $L$ now perform closely to the conventional single codebook precoding that has a codebook $\mathcal{F}^{L+\log K}$ with $L + \log K$ feedback bits,

$$\mathbf{F}_{\hat{i}} = \arg\max_{\mathbf{F}_i \in \mathcal{F}^{L+\log K}} C(\mathbf{F}_i), \tag{15}$$

in the sense that the rotating codebooks $\{\mathcal{F}_k^L, 0 \leq k \leq K-1\}$ eventually select the same optimal precoding matrix as the single codebook $\mathcal{F}^{L+\log K}$ as a result of the use of the default precoding matrix with certain delay. That is a gain of $\log K$ bits without actually committing $\log K$ additional feedback bits.

However, it requires more selection time for the rotating codebook of feedback bits $L$ to select the optimal precoding matrix from $K2^L = 2^{L+\log K}$ matrices from $K$ codebooks of size $2^L$ than for the single codebook of $L + \log K$ feedback bits to select the same optimal matrix from $2^{L+\log K}$ matrices from a single codebook. For the single codebook with feedback bits of $L + \log K$, the selection of the optimal matrix is instantaneous since with $L + \log K$ feedback bits, all $2^{L+\log_2 K}$ matrices are immediately accessible at each feedback time. While for the rotating codebook with $L$ feedback bits, the $2^{L+\log K}$ matrices are spread over $K$ codebooks and only $2^L$ matrices are accessible at each feedback instance. The rotating codebook may (in the worst case) take a full rotation period ($K$ feedback intervals) to obtain the optimal precoding matrix provided by a single codebook $\mathcal{F}^{L+\log K}$. Prior to that happening, only a suboptimal matrix can be chosen from the rotated codebooks at each feedback. With each rotation, more codebooks are traversed; a better suboptimal matrix is hence selected at each feedback. After $K$ rotations ($K$ feedback intervals, $KT$), all codebooks are searched and the optimal matrix is guaranteed to be selected. The average delay is $\frac{1}{2}KT$. Therefore, the rotating codebook precoding with $L$ feedback bits and $K$ rotating codebooks essentially performs the same as the conventional precoding with feedback bits of $L+\log K$ when the channel coherence time $T_c$ is much longer than $\frac{1}{2}KT$,

$$T_c \gg \frac{1}{2}KT. \tag{16}$$

That is, the effective codebook size of the rotating codebooks with feedback bits $L$ is $K2^L$ which is $K$ times larger than the conventional codebook without committing $\log_2 K$ extra feedback bits.

As the channel coherence time becomes shorter, condition (16) no longer holds. The rotation delay becomes a significant part of the coherence time. Before the optimal precoding matrix is selected, only sub-optimal precoding matrices can be selected during the delay period. For example, after the first codebook rotation, the best precoding matrix from two codebooks (the current one and the previous one) is selected for current precoding. After the second rotation, the best precoding matrix is selected from three codebooks (the current one and the previous two). After the $K$-1th rotation, the best precoding matrix can then be chosen from the $K$ codebooks. The *effective* codebook size $\kappa 2^L$ is hence larger than $2^L$ but less than $K2^L$, i.e., $1 < \kappa < K$, corresponding to $L + \log_2 \kappa$ effective feedback bits. As the channel coherence time becomes shorter than the rotation period, i.e., $T_c < KT$, only $T_c/T$ codebooks can be rotated within the coherence time. The effective number of codebooks $\kappa$ becomes $1 < \kappa < T_c/T$. The corresponding effective feedback bits is $L + \log_2 \kappa$.

In general,

$$1 \leq \kappa \leq \min\{T_c/T, K\} \tag{17}$$

i.e., the effective number of feedback bits $L' = L + \log_2 \kappa$ is

$$L \leq L' \leq L + \log_2 K \tag{18}$$

That is, the rotating codebook performance is upper bounded by (15) and lower bounded by (7). In other words, the rotating codebook precoding performance with $L$ feedback bits and $K$ rotating codebooks is upper bounded by the conventional precoding performance with $L+\log K$ feedback bits (when $T_c \gg KT$) and lower bounded by $L$ feedback bits (when $T_c < T$). The rotating codebook gain over the conventional precoding with the same feedback bits is thus always non-negative.

The $K$ codebooks used for rotation is created by generating $K$ independent but equivalent codebooks. One can also create such $K$ codebooks by simply splitting a larger codebook (mother codebook) $\mathcal{F}^{L'}$ into $K=2^{L'-L}$ equivalent smaller codebooks (child codebooks) $\{\mathcal{F}_k^L, \ 0 \leq k \leq K-1\}$, each of size $2^L$, where $\bigcup_{k=0}^{K-1} \mathcal{F}_k^L = \mathcal{F}^{L'}$ and $L' > L$. The remainder of the paper will utilize the latter approach for creating rotating codebooks.

As a result of the use of the reserved index for indication of the default precoding matrix, total of $K$ matrices from the mother codebook (one from each of the child codebook) will not be available for use by the transceivers. To avoid the precoding matrix $\mathcal{F}_k^L(\tilde{i}), k=0,\cdots,K-1$, corresponding to the reserved index $\tilde{i}$, from being permanently punctured from the codebook $\mathcal{F}_k^L$, the reserved index $\tilde{i}$ should be made variable over time. That is, instead of fixing $\tilde{i}$, $\tilde{i}$ is now made time variant according to a predetermined pseudo random sequence $0 \leq \varsigma(t) \leq 2^L - 1$ known to both the transmitter and the receiver. That is, $\tilde{i} \triangleq \varsigma(t)$ at time $t$. Fig. 4 illustrates this random puncturing approach. It is evident that the precoding matrix being punctured changes from time to time. In effect, the precoding matrix $\mathcal{F}_{\rho(t)}^L(\varsigma(t))$ punctured from codebook $\mathcal{F}_{\kappa(t)}^L$ at feedback time $t$ will be available at feedback time $t+KT$ when a different matrix $\mathcal{F}_{\rho(t+KT)}^L(\varsigma(t+KT))$ is punctured. Therefore, all the precoding matrices from the mother codebook are virtually available over time for the transceivers.

## IV. SIMULATION RESULTS

Simulations were performed to verify the rotating codebook performance using a MIMO-OFDM simulator. Fig. 5 shows the performance gain of the rotating codebooks with $L=4$ and $K=2^6$ (i.e., the number of feedback bits is 4 and the number of rotating codebooks (each of size $2^4$) is $2^6 = 64$) over the conventional precoding with various feedback bits. The method for creating the codebooks is the same as in Section II. It is clear that, at 3 km/h, the rotating codebook scheme improves the precoding performance from the conventional precoding with $L=4$ (lower bound for $L=4$, $K=2^6$) to better than that of the conventional precoding with $L=8$ feedback bits, corresponding to an effective gain of $\log_2 \kappa = 8-4 = 4$ feedback bits (and effectively $\kappa = 2^4 = 16$ times larger codebook) converting to 2 dB gain in SNR. In fact, for further lower mobility, $L=4$ with $K=2^6$ can approach closer to the performance of $L=10$ (upper bound for $L=4$, $K=2^6$) as shown in Fig. 6 where the mobile speed is 1 km/h.

On the other hand, as the mobility increases, the rotating codebook gain over the conventional precoding is reduced accordingly. Fig. 7 shows the rotating codebook precoding performance at 15 km/h with decreased gain. The rotating codebook gain over the conventional precoding is less but still significantly better than conventional $L=4$. Fig. 7 also shows that the rotating codebook gain diminishes at 60 km/h where the rotating codebook precoding with $L=4$ and $K=2^6$ performs the same as the conventional precoding with $L$=4 (lower bound), although at high mobility the closed-loop precoding performance difference among different values of $L$ is already small.

It is worth noting that, in theory, the static (i.e., $T_c = \infty$) performance of the rotating codebook precoding with $L=4$ approaches $L=\infty$ as $K \to \infty$. However, in a practical mobile environment, the rotating codebook gain is ultimately capped by the channel coherence time. The choice of $K$ therefore depends on the lowest mobility that the application is targeted at to reduce unnecessary memory for storage of codebooks. For environment with mobility higher than 1 km/h, $K = 2^6 = 64$ is obviously more than sufficient.

## V. Conclusion

The full gain from MIMO precoding is achieved with full CSI at the transmitter since this allows the transmitted signal to be shaped based on the eigen-structure of the channel. Feedback of CSI to the transmitter can thus enable the transmitter to better exploit channel conditions to improve the MIMO performance. However, the amount of channel information fed back to the transmitter (i.e., the size of the codebook) is limited by the often severely limited feedback control channel bandwidth, thereby preventing the transmitter from obtaining full channel information in order to fully exploit MIMO precoding gain. In this letter, presented is a rotating codebook precoding approach that virtually extends the size of the codebook without the need of increasing the feedback bits. That is, the rotating codebook provides the transceivers with essentially a larger codebook, while utilizing the same number of feedback bits of a smaller codebook. This is made possible by the use of two key mechanisms: 1) multiple codebook rotation that allows transceivers to see more precoding matrices without utilizing more feedback bits; and 2) default precoding matrix that enables the combining of the optimization over previous codebooks with the current codebook. As a result, compared to the conventional closed-loop precoding scheme, the transceivers have more precoding matrices for optimization, improving the closed-loop precoding performance. The actual gain of the rotating codebook over the conventional closed-loop precoding is determined by the effective number of codebooks that the transceivers can traverse within the channel coherence time. The rotating codebook precoding hence provides more gain over the conventional closed-loop precoding with the same feedback bits in stationary and pedestrian environment (which is typically the scenario that high rate data applications is targeted at) and reduces to the same performance as the conventional closed-loop precoding at high mobility. That is, the rotating codebook precoding automatically and fully exploits the channel coherence provided by the environment and converts it to the performance advantage, which is not the case for the conventional precoding. Although this letter uses unitary precoding as the application paradigm, the rotating codebook scheme is applicable to any codebook based precoding system with limited feedback.

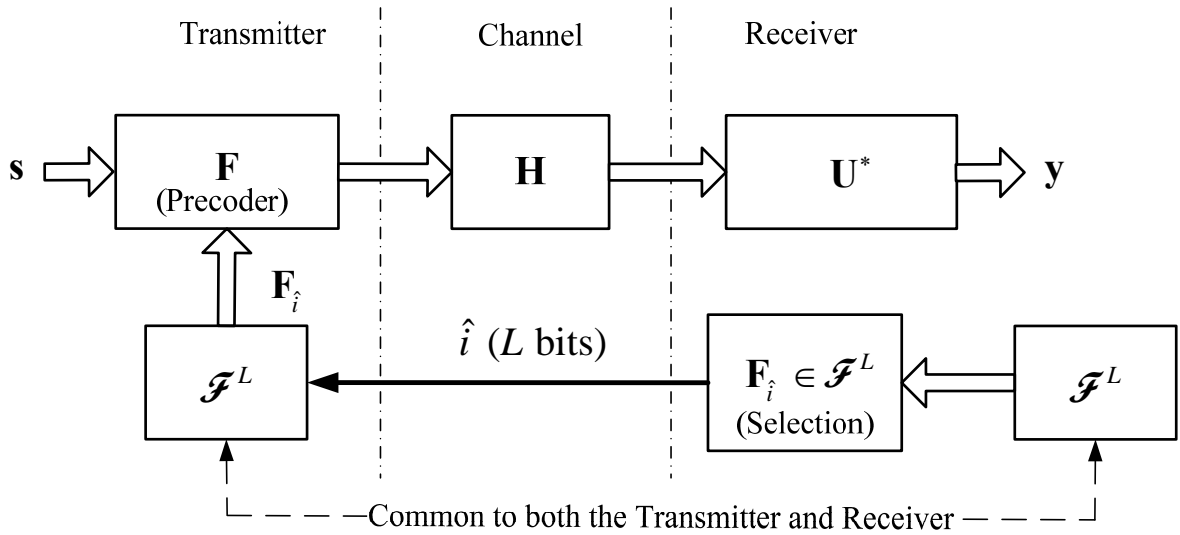

Fig. 1. Block diagram of a limited feedback unitary precoding MIMO system.

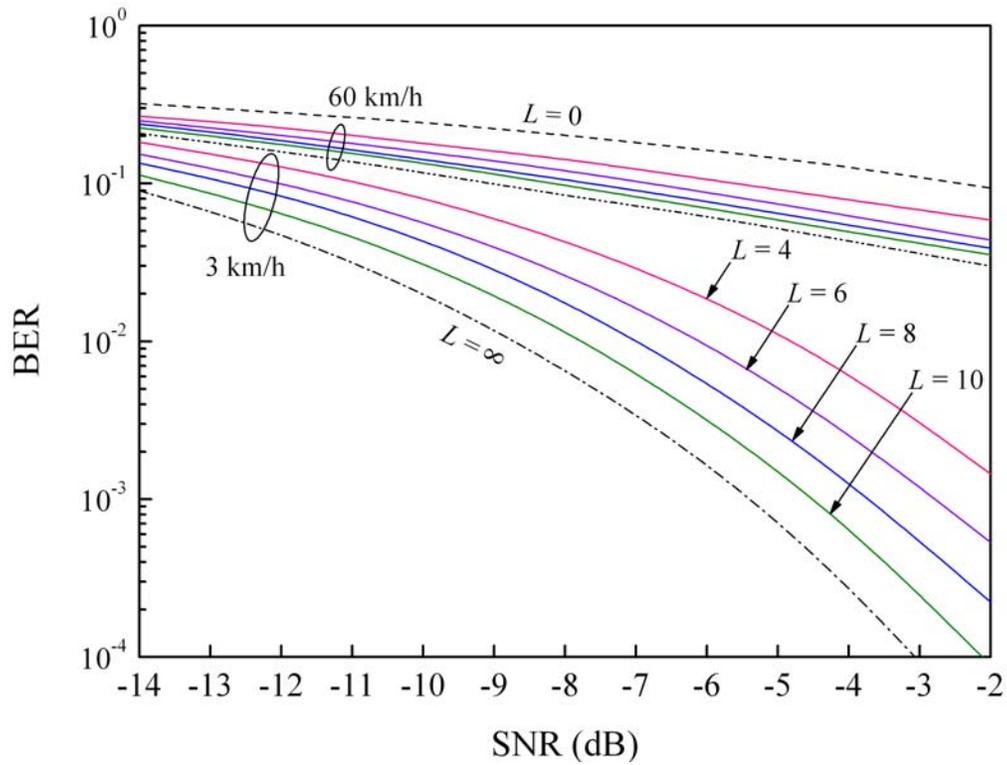

Fig. 2 Unitary precoding performance for $M_T = 8$, $M_R = 1$ antenna configuration. Fading speeds = 3 km/h and 60 km/h at 2GHz carrier frequency; Number of feedback bits = 0 (no feedback), 4, 6, 8, 10, and $\infty$; Feedback interval=3 msec.

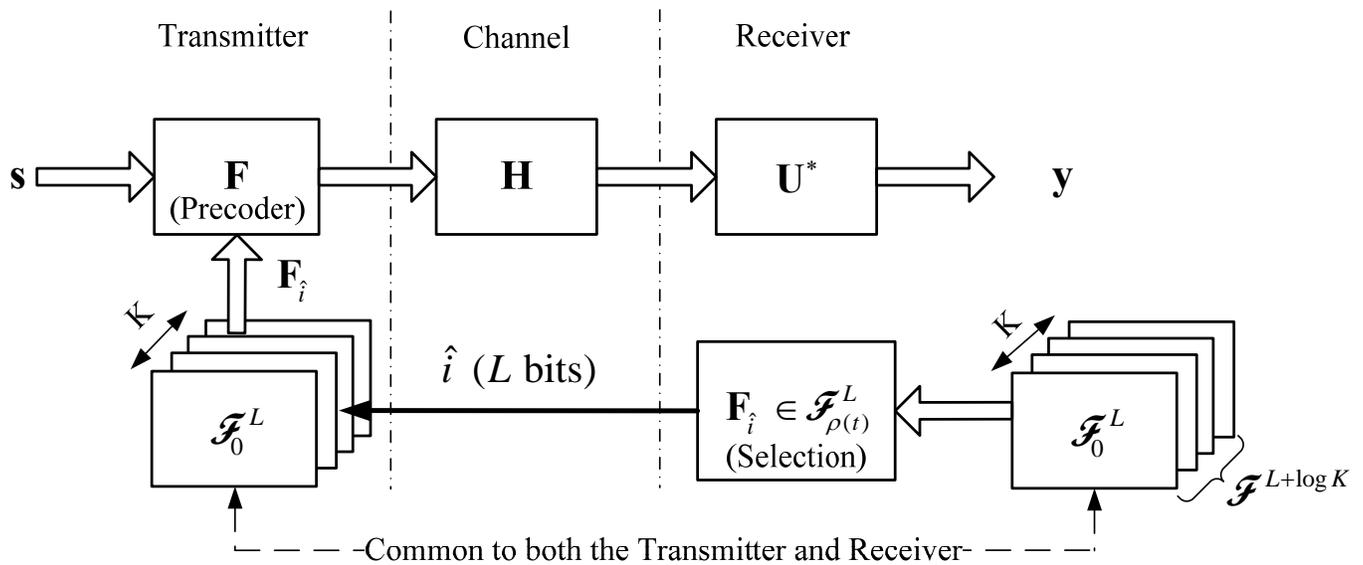

Fig. 3. Schematic block diagram of the rotating codebook precoding system where the codebook used at feedback time $t$ is $\mathscr{F}_{\rho(t)}^{L}$.

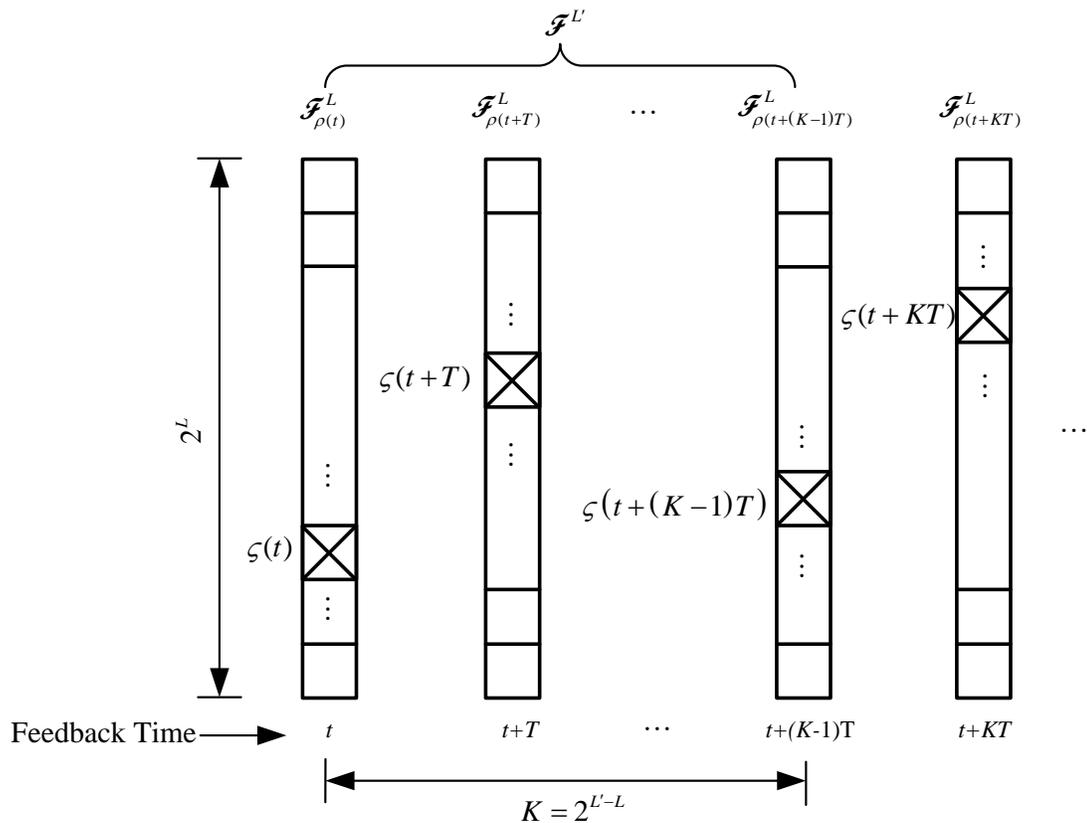

Fig. 4. Illustration of codebook random puncturing for rotating codebook precoding, where "×" denotes the precoding matrix that is punctured and replaced by the default precoding matrix.

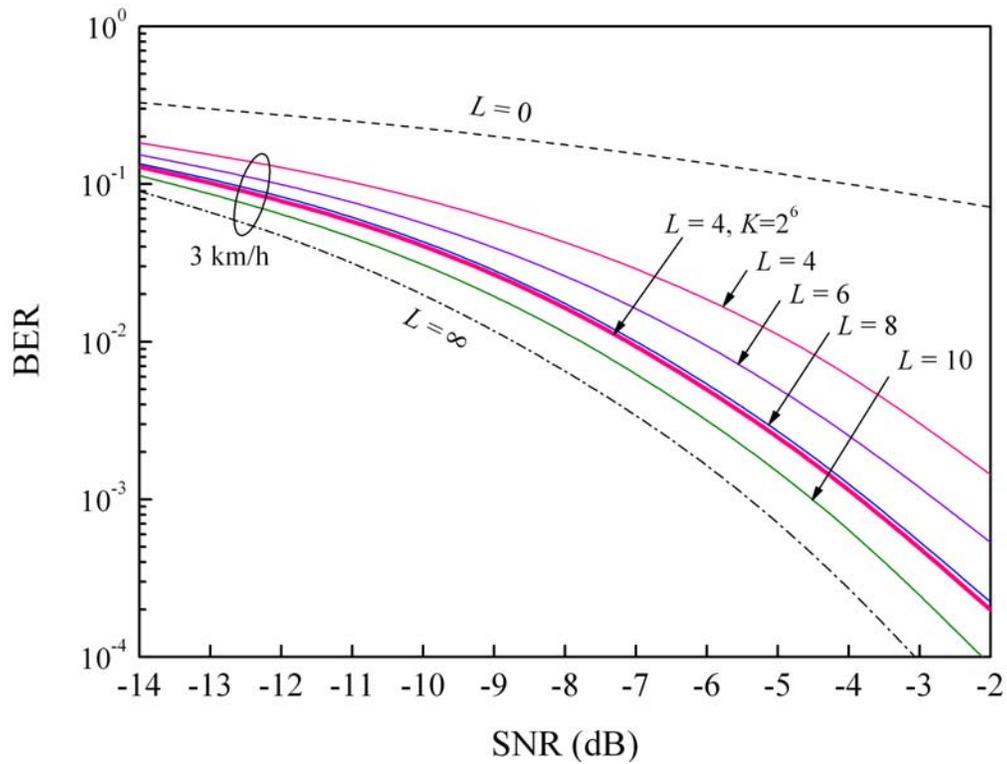

Fig. 5 Rotating codebook performance with $L = 4$ and $K = 2^6$ ($M_T = 8, M_R = 1$) at 3 km/h/2 GHz carrier frequency. The number of feedback bits for the conventional precoding is $L = 4, 6, 8,$ and $10$; Feedback interval = 3 msec.

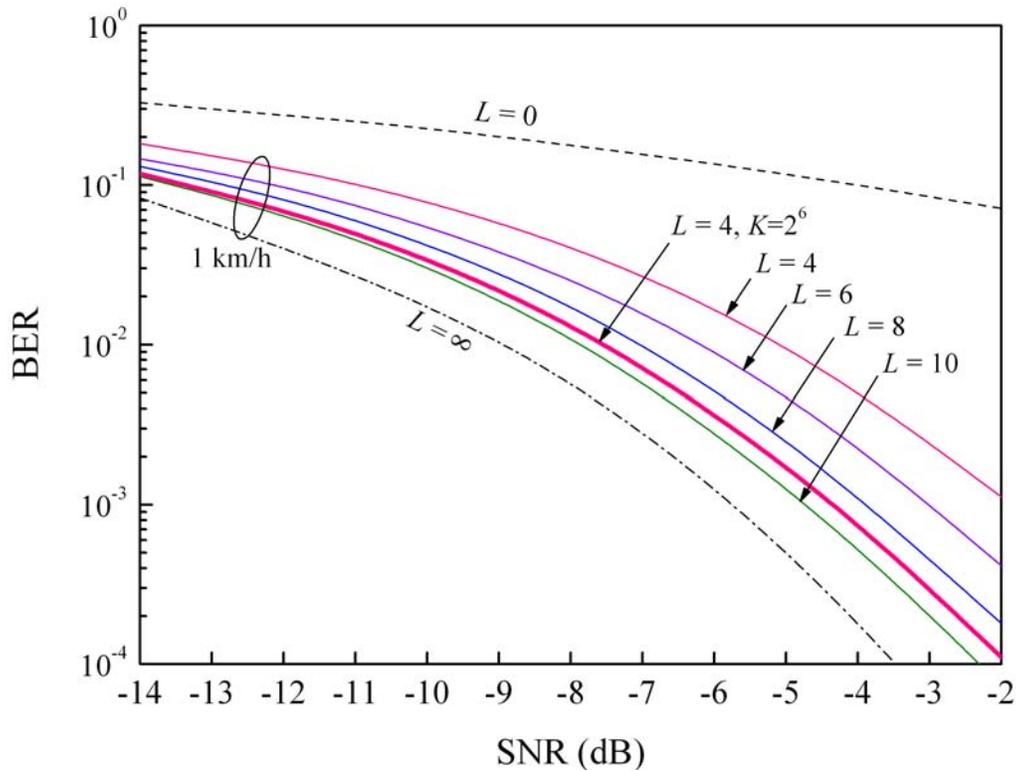

Fig. 6. Rotating codebook performance with $L = 4$ and $K = 2^6$ ($M_T = 8, M_R = 1$) at 1 km/h at 2 GHz carrier frequency. The number of feedback bits for the conventional precoding is $L = 4, 6, 8,$ and $10$; Feedback interval = 3 msec.

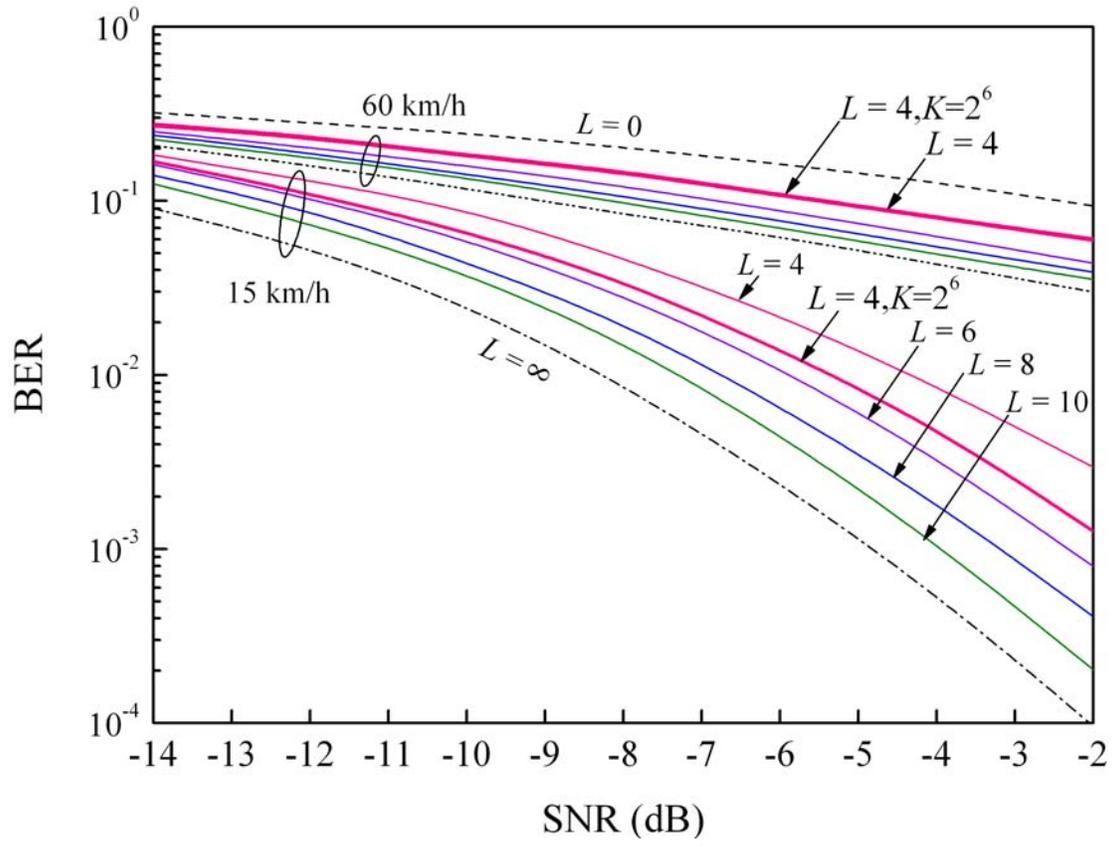

Fig. 7 Rotating codebook performance with $L=4$ and $K=2^6$ at 15 and 60 km/h ($M_T=8, M_R=1$).